\documentclass[manuscript]{aastex}
\newcommand{\um}{\,\mu {\rm m} }
\newcommand{\mJySr}{\,{\rm MJy/Sr} }
\newcommand{\mJyBm}{\,{\rm mJy/Bm} }
\newcommand{\mK}{\,{\rm mK} }

\newcommand{\uJy}{\,{\rm \mu Jy} }
\newcommand{\uK}{\,{\rm \mu K} }

\newcommand{\ghz}{{\, \rm GHz}}

\begin{document} 

\title{A Limit on the Polarized Anomalous Microwave Emission of Lynds
1622}

\author{B. S. Mason}
\affil{National Radio Astronomy Observatory, 520 Edgemont Rd. Charlottesville VA 22904}
\email{bmason@nrao.edu}
\author{T. Robishaw\altaffilmark{1}}
\author{C Heiles}
\affil{University of California, Berkeley, 601 Campbell Hall Berkeley CA 94720}
\author{D. Finkbeiner}
\affil{Harvard-Smithsonian Center for Astrophysics, MS 51 61 Garden St. Cambridge, MA 02138}
\author{C. Dickinson} 
\affil{Infrrrared Processing and Analysis
Center, California Institute of Technology, MS 220-6, 1200
E. California Blvd., Pasadena, CA 91125}
\altaffiltext{1}{current address: 570 Physics Annex, University of Sydney School of Physics, NSW 2006 Australia}

%
%
%
%
%
%
%

\begin{abstract}

The dark cloud Lynds 1622 is one of a few specific sites in the Galaxy
where, relative to observed free-free and vibrational dust emission,
there is a clear excess of microwave emission. In order to constrain
models for this microwave emission, and to better establish the
contribution which it might make to ongoing and near-future microwave
background polarization experiments, we have used the Green Bank
Telescope to search for linear polarization at $9.65$ Ghz towards
Lynds 1622. We place a $95.4\%$ upper limit of $88 \uK$ ($123 \uK$ at
$99.7\%$ confidence) on the total linear polarization of this source
averaged over a $1'.3$ FWHM beam.  Relative to the observed level of
anomalous emission in Stokes I these limits correspond to fractional
linear polarizations of $2.7\%$ and $3.5\%$.

\end{abstract}


\section{Introduction}

The discovery of anomalous dust-correlated emission by multi-frequency
Microwave Background experiments in the 1990s \citep{leitch97,
deoliveiracosta97,kogut96} led to a recalculation by \citet{draine98a}
of the electric dipole emission from small rotating dust grains, first
considered by \citet{erickson57}.  The emission is spatially
correlated with dust, but far in excess of a reasonable extrapolation
of the thermal emission \citep{fink99} and is therefore called
``anomalous'' dust emission, or ``Foreground X''
\citep{deoliveiracosta02} because of its interference with cosmic
microwave background (CMB) experiments.  Since the \citet{draine98a}
paper on spinning dust, and a later paper proposing magnetic dipole
emission from Fe containing grains \citep{draine99} measurements of
this component have been refined
\citep[e.g.][]{deoliveiracosta98,fink02,banday03,deoliveiracosta04,fink04,davies06}
using various data sets.

One yet untested prediction of the spinning dust theory is that the
emission should be moderately polarized ($3-7\%$) at low frequencies,
falling to under 2\% above 20 GHz \citep{lazarian00}. At 10 GHz 3-5\%
linear polarization is expected. In contrast magnetic dipole
fluctuations in larger single-domain grains would show a strong linear
polarization ($\sim 10\%$) below 10 GHz, rising to over 30\% at 100
GHz \citep{draine99}, although the details depend on the shape, iron
fraction, and magnetic domain configuration of the grains.  Because
the polarization properties of the anomalous dust emission are
unknown and are of immense interest to the CMB community, a deep
exposure on a cloud known to contain the anomalous emission would make
a substantial contribution to our understanding of this component.

Although anomalous, dust-correlated microwave emission has been
detected statistically by numerous experiments, at the time of writing
the number of {\it individual} lines of sight on which it is seen are
but three: the NCP loop \citep{leitch97}; Lynds 1622
\citep{fink02,casassus04}; and Perseus
\citep{watson05}. \citet{battistelli06} report the only measurement of
the polarization of the anomalous emission, a detection of 3\%
fractional polarization in the Perseus cloud with the COSMOSOMAS
experiment.  We have used the GBT at $9 \, {\rm GHz}$ to search
for polarization towards Lynds 1622 (L1622).

\section{Observations \& Calibration}

We chose to search for polarized emission from L1622 along the line of
sight 05:54:23, $+01$:46:54 (J2000); this is the same pointing
position used by Finkbeiner et al. in previous 9 GHz measurements of
L1622. It is also near the peak of 30 GHz excess emission observed by
CBI \citep{casassus06}.  We observe at $9 \, {\rm GHz}$, a frequency
where the product of fractional linear polarization and total
intensity for spinning dust is expected to be detectable (a few
mK). This is also the lowest frequency GBT receiver equipped with
circularly polarized feeds, which are essential to obtaining the
broadband continuum stability needed for the Stokes Q and U
measurements. Two sub-bands were chosen by examining site RFI monitor
data, each of 200 MHz total bandwidth, centered on $8.65$ and $9.65$
GHz, respectively.  Much wider bandwidth measurements are in principle
possible, but RFI was a key consideration. The receiver temperature is
smooth and low in these regions and avoid known spectral resonances.

The GBT X-band receiver has a single, dual-circularly polarized feed
horn.  The GBT Spectrometer was used to form all four auto- and
cross-correlations between the IF signals from each of LCP and RCP,
providing in principle instantaneous measurements of all four Stokes
parameters ($I,Q,U,V$). The Stokes parameters describing linear
polarization $(Q,U)$ principally comprise combinations of the
cross-correlations, which has the advantage that receiver gain
fluctuations and atmospheric emission variations are suppressed.

Successive pairs of On/Off measurements on L1622 were performed, each
lasting 48 seconds and consisting of 1-second integrations.  The
On/Off pairs were matched in hour angle to provide approximately the
same track in azimuth and elevation for each pair; allowing for
scan-related observing overheads we found we required a separation of
$\sim 66$ seconds in Right Ascension.  This strategy minimizes the
potential effect of polarized ground spillover, which would be
expected to have a signature that changes with azimuth and
elevation. Note however that the expected level of spillover is low
due to the clear aperture, off-axis design of the GBT; deep
integrations with the GBT $26 - 40 \, {\rm GHz}$ receiver show that
the level of any systematic signal in total intensity is $< 100 \,
\uJy$.  The choice of trailing field was constrained by a compromise
between observing efficiency and the benefits of short cycle times,
particularly for the total intensity measurement.  The timing accuracy
of the GBT control system was not sufficient to allow an analogous
LEAD region.  In all ~700 On/Off pairs were collected in 30 hours of
observing. Every hour the telescope pointing and focus is checked on
the nearby calibrator 3C138. 3C138 is also a well-measured
polarization calibrator and we perform an On/Off measurement of 3C138
with the system configured identically to our L1622 polarization
observations.

Prior to any calibration the data are flagged for RFI. Several fixed
frequency ranges show common interfence events and are flagged in all
scans ($9.57579 - 9.5777$ GHz; $9.6673 - 9.6688$ GHz; $9.7005 -
9.7018$ GHz; and any data over $9.730$ GHz). Beyond this the RL and LR
data for each IF are searched and integration with a $>5\sigma$ spectral
feature is flagged for rejection. 

The full-Stokes data were calibrated using a variation on the
procedures described in 
\citet{heiles01a,heiles01b,heiles01c}, which we briefly
review. A single noise diode is used to inject noise into each of LCP
and RCP, and this (coherent) signal is used to determine the phase
difference between the two IF paths. The dominant contributor to the
R/L phase is the path length difference between the signal chains
($\sim 70 \, {\rm cm}$).  The response of the instrument to celestial
polarization is described by the Mueller matrix (MM).  We determine
the MM from the 3C138 measurements assuming a fractional linear
polarization of $11\%$ and a parallactic angle of $170^{\circ}$ of
the linear polarization pseudovector.  The data are also
amplitude-calibrated using 3C138, for which we assume flux densities
$S_{8.65} = 2.43 \, {\rm Jy}$ and $S_{9.65} = 2.20 \, {\rm Jy}$.  This
allows the determination of an effective Stokes I flux density of the
calibration diode, $S_{cal}$, as a function of frequency across our
observing band. After correcting for the measured LR phase difference
and calibrating data for the bandpass as a function of frequency from
the $S_{cal}$ data, means across the band are taken to form nominal
Stokes $(I,Q,U,V)$ measurements. Mueller Matrices are computed from
these data on the calibrator, and applied to the equivalent data on
L1622.  The result of these procedures is a position-switched Stokes
$(I,Q,U,V)$ in Janskys per beam for each of 720 ON-OFF measurements of
L1622 in our dataset.  This is converted into a main-beam filling
equivalent surface brightness by multiplying by the GBT gain (taken to
be $1.95 \, {\rm K/Jy}$ at $9.65 \ghz$) and dividing by the main beam
efficiency ($\eta_B \equiv \Omega_{mb}/\Omega_{ant} = 93\%$). 


The beam properties were determined by scans along six evenly spaced
directions through 3C286, and orthogonal (az/el) scans through
3C138. We adopt a beam width of $1'.30 \, ({\rm FWHM})$. 


\begin{figure*}[h!]
\plotone{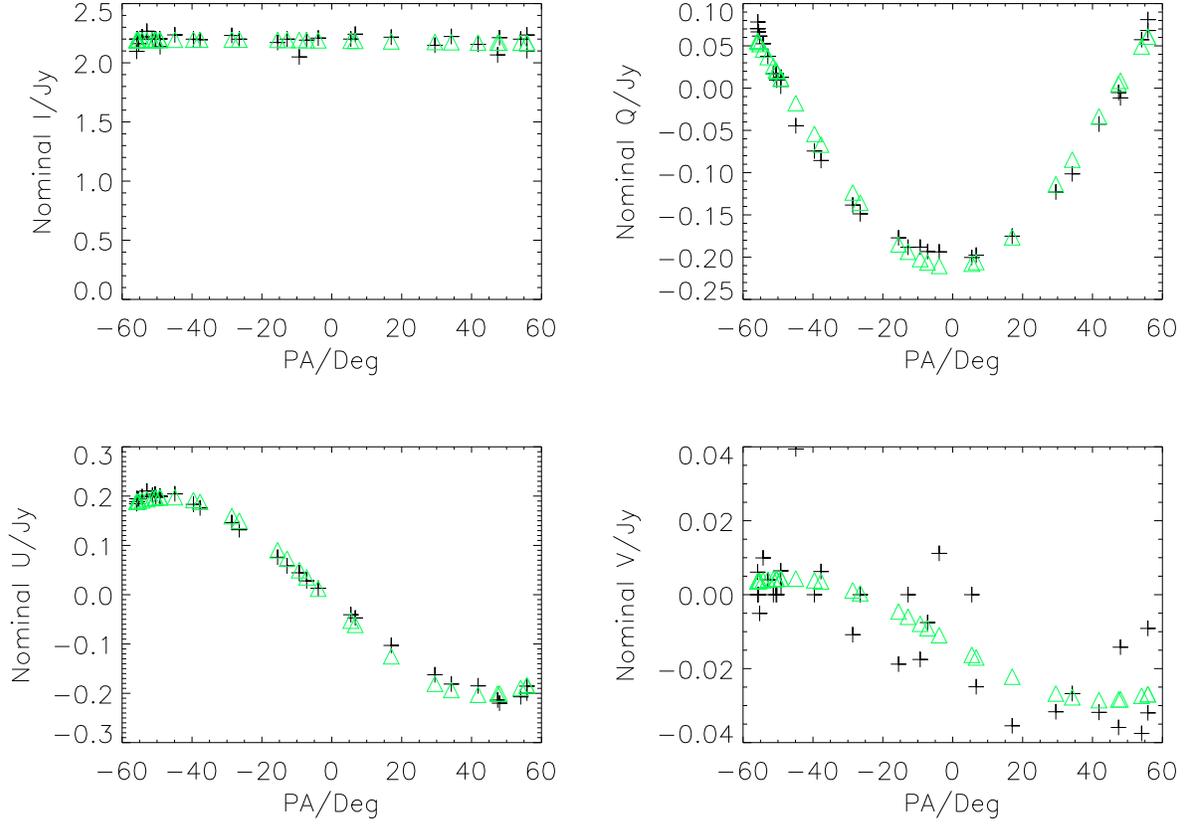}
\caption{Measurements of 3C138 (black pluses) and the fitted Mueller
  Matrix describing the GBT polarization response (green
  triangles). The quantities shown are, from top left, the average
  autocorrelation (LL+RR); the real part of the LR cross correlation;
  the imaginary part of the LR cross correlation; the the difference
  between the autocorrelations.}
\label{fig:calib}
\end{figure*}

The noise level was estimated from the data for each measurement by a
robust, iterative procedure. First the the median absolute
deviation\footnote{The median absolute deviation (MAD) is defined as
$Median(|x_i - Median(x_i)|)$ and is much less sensitive to outliers
than the variance, although its sampling variance is greater.  For a
Gaussian distribution $MAD = 0.674 \sigma$; in data analysis, MAD
values are normalized to a Gaussian-equivalent $\sigma$.} of the
measurements is computed for a given IF and Stokes parameter in a
3-hour buffer centered on the time of the measurement. We refer to the
scatter so computed in this first pass on the data as $\sigma_1$.
Data more than $6\sigma_1$ from the mean are rejected iteratively,
{\it i.e.}, if a data point(s) over the threshold exists, the worst
outlier is rejected, the $\sigma$ recomputed, and the process repeated
until there are no data over the threshold.  This rejects $\sim 2\%$
of the data. After this process is completed the noises are recomputed
using the variance in a sliding 3-hour buffer, resulting in a noise
estimate $\sigma_2$ on the outlier-rejected data. Data with $\sigma_2$
values more than 5 times the expected radiometer noise ($S_{max,QU}= 5
\times 0.2 \, \, {\rm mJy}$) in $Q$ and $U$ are rejected. Stokes $I$
data with noises more than $5$ times the {\it best} RMS ($S_{max,I}=5
\times 10 \, {\rm mJy}$) are rejected. This removes data collected in
periods of less than optimal weather. Final results are computed as
the weighted mean of individual measurements, with weights of
$\sigma^{-2}_{2}$. The final results for each polarization are shown
in Table~\ref{tbl:finalresults} along with the average noise levels.
Results for the $8.65$ GHz channel are consistent with those from the
$9.65$ GHz channel but the Stokes $Q$ and $U$ parameters show noise
levels a factor of $\sim 2$ higher, perhaps reflecting a hardware
instability in these channels; we exclude them from our final results.
The large variation in noise levels between the Stokes $I$, $Q$, $U$,
and $V$ results is explained by the fact that with our observing
technique both Stokes $I$ and $V$ are affected by receiver gain
fluctuations, and Stokes $I$ is additionally affected by fluctuations
in atmospheric emission, whereas Stokes $Q$ and $U$ largely are not.

\begin{figure*}[h!]
\plotone{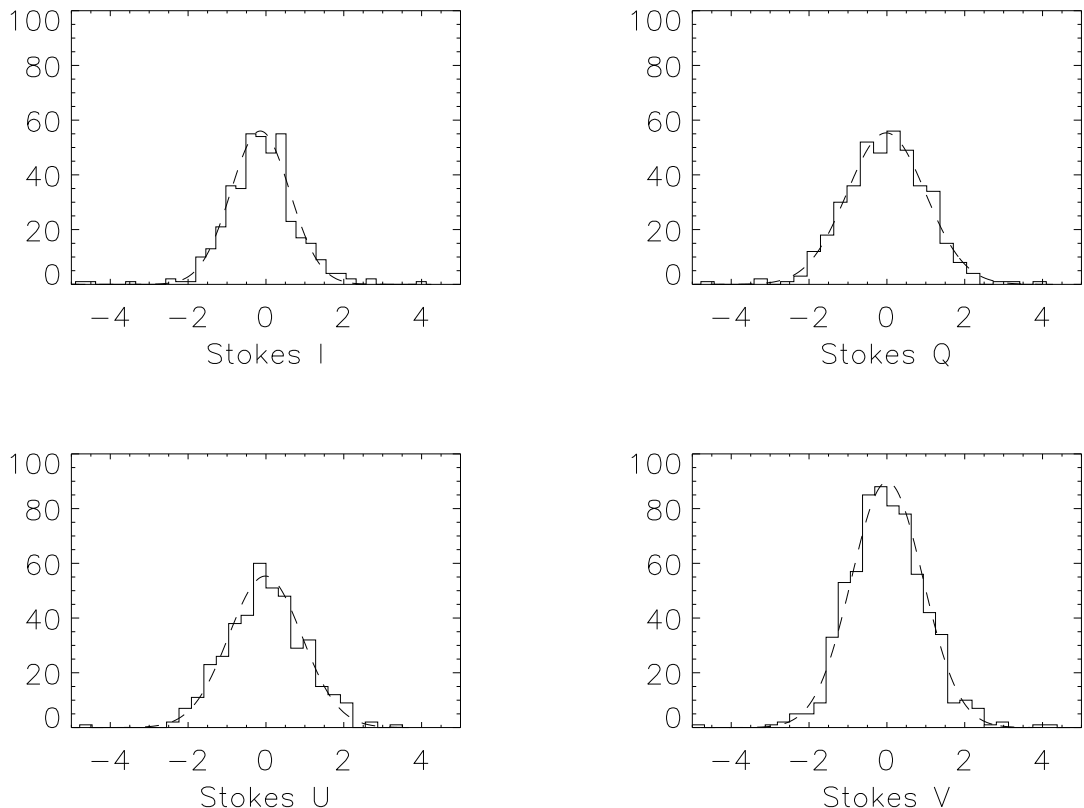}
\caption{The distribution of individual measurements divided by their
noise estimates ($8.65$ GHz is black and $9.65$ GHz is magenta). The
dashed lines show the best-fitting Gaussian to each distribution. In
all cases the Gaussian has a $\sigma$ within $4\%$ of unity indicating
the accuracy of our noise estimate.}
\label{fig:tdist}
\end{figure*}

\begin{figure*}[h!]
\plotone{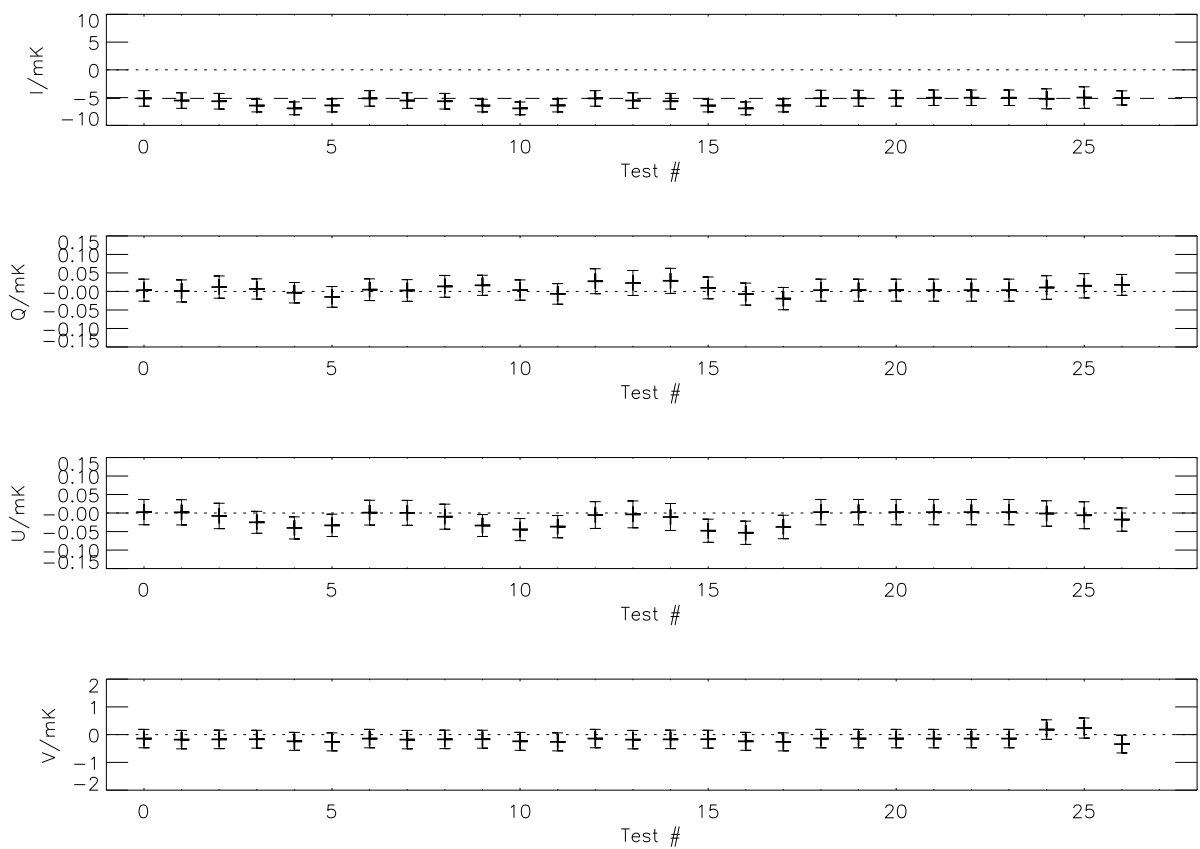}
\caption{Fully averaged Stokes $I$, $Q$, $U$, and $V$ results for a
range of filtering parameters.  The dashed line shows our adopted
value (test 0). }
\label{fig:datatests}
\end{figure*}

A histogram of the individual measurements, normalized by the
$\sigma_2$ value for each measurement is shown in
Figure~\ref{fig:tdist} along with the best-fit Gaussian to the
distribution which should have a Gaussian $\sigma$ close to unity if
our noise estimate is accurate.  For Stokes $Q$ and $U$ the fitted
Gaussians are within $5\%$ of unity; for $I$ and $V$, within $12\%$ of
unity.  While the data on the whole are well-described by a Gaussian
distribution there are a small number of outliers. To determine the
sensitivity of our results to these outliers, and to the noise
estimate that results from our pipeline, we have varied our data
filtering parameters in a suite of tests summarized in
Table~\ref{tbl:systests}.  The controlling parameters of our filtering
procedure were systematically varied; the fully-averaged $9.65$ GHz
Stokes parameters that result for each variation are shown in
Figure~\ref{fig:datatests}. Note that our adopted parameters
are denoted as ``Test 0''.

\begin{figure*}[h!]
\plotone{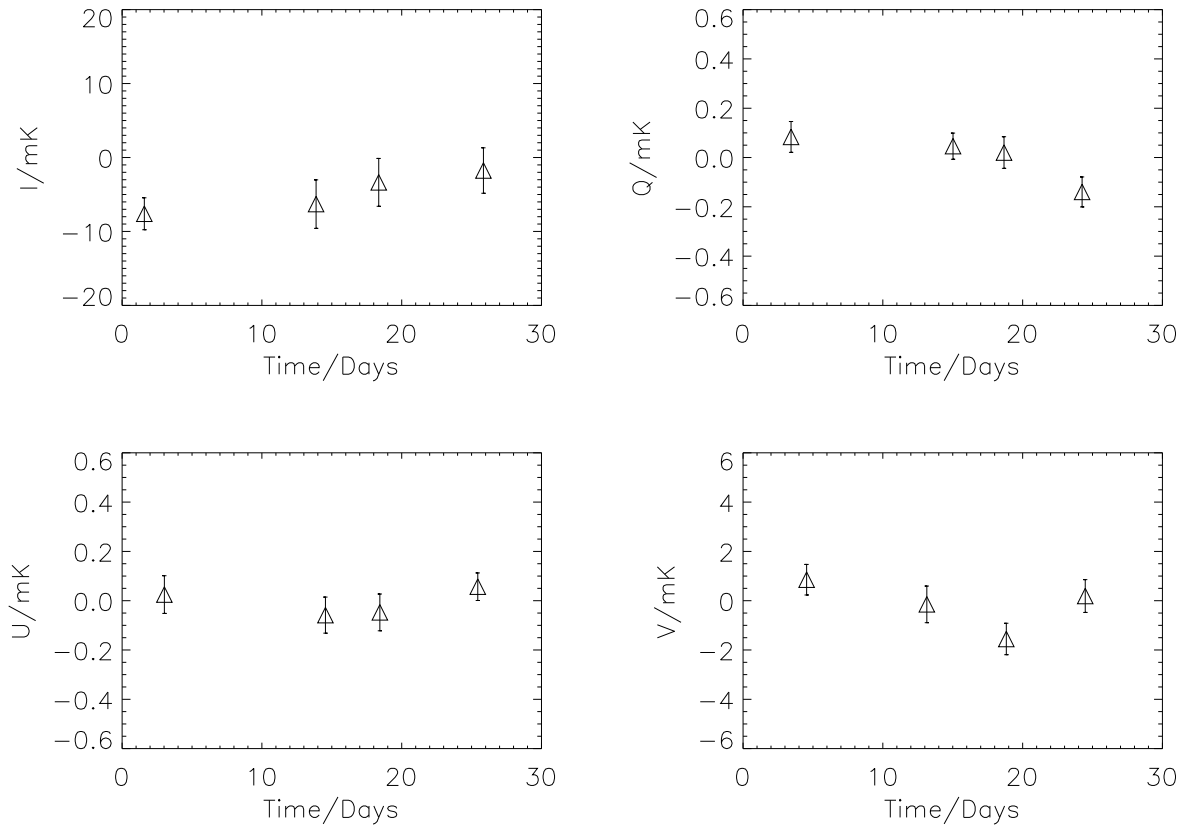}
\caption{Mean values for each of Stokes $I$,$Q$,$U$, and $V$ with data binned in time over
the course of the observing run.}
\label{fig:Stokesmjd}
\end{figure*}

\begin{figure*}[h!]
\plotone{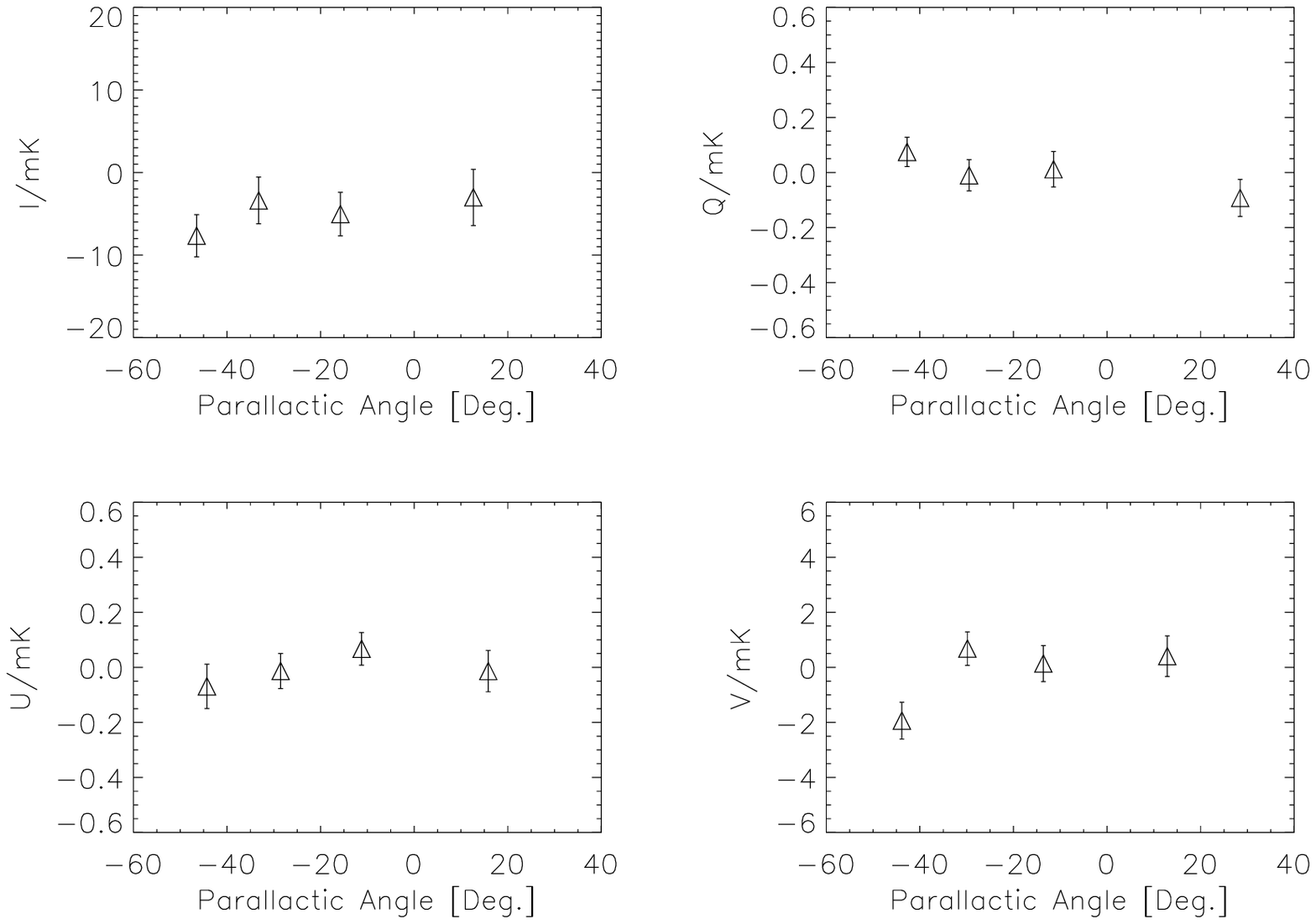}
\caption{Mean values for each of Stokes $I$,$Q$,$U$, and $V$ with data binned in parallactic angle.}
\label{fig:Stokespa}
\end{figure*}

The data binned by time and Parallactic Angle (PA) are shown in
Figures~\ref{fig:Stokesmjd} and \ref{fig:Stokespa}, respectively.  The
overall consistency of the data are good, with the most significant
deviation in Stokes $V$. The absence of significant variations in
Stokes I, Q, and U with PA indicates that any residual ground signal
is lower than the sensitivity achieved.

\begin{figure*}[h!]
\plotone{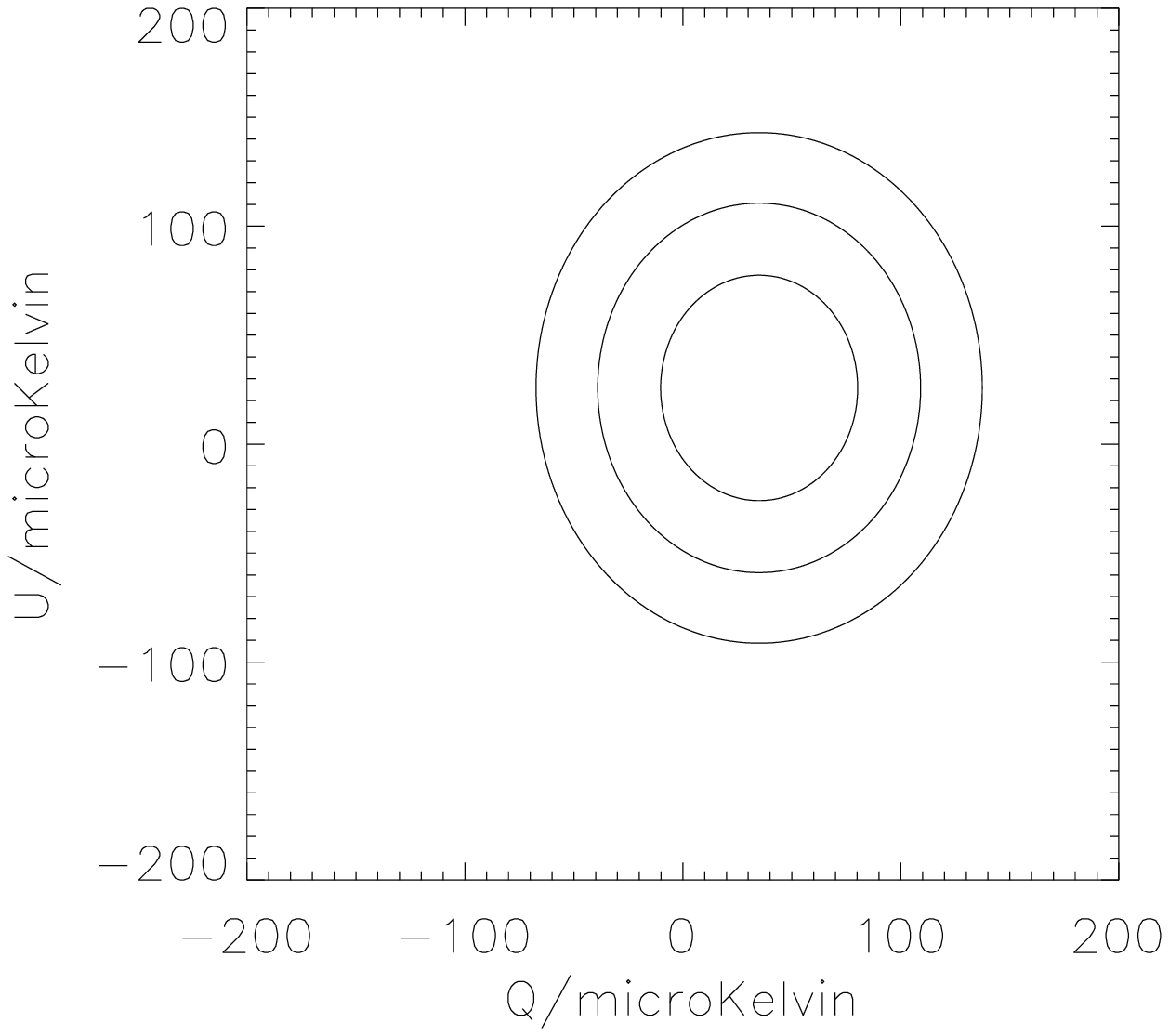}
\caption{Joint $68.3\%$, $95.4\%$, and $99.7\%$ confidence regions for Stokes Q and U.}
\label{fig:qulimits}
\end{figure*}

Joint $1\sigma$, $2\sigma$, and $3\sigma$ confidence intervals on the
$9.65$ GHz linear polarization signal are shown in
Figure~\ref{fig:qulimits}. These represent the difference in Stokes
$Q$ and $U$ between the ON and OFF source locations, and on the
assumption that the background signal is unpolarized at our observing
frequency are limits on the polarization on the line of sight to
L1622.  

We note that there is a {\it negative} signal in Stokes $I$ on the
line of sight to L1622. This is a robust result, likely due to
small-scale variations in the free-free signal in the region; we
discuss this possibility further in \S~\ref{sec:Stokesiinterp}.

\begin{table*}[h]
\begin{center}
\begin{tabular}{l l l l l l l }
Test        & Buffer 1       & Buffer 1 Reject           & Buffer 2         & Buffer 2 Reject                & $Q,U$ Max.                       & $I$ Max.                  \\
Number      & Scatter        &  Threshold                 &   Scatter        & Threshold                     & Buffer Noise                     &  Buffer Noise             \\
            & Method         & ($S_{cut,1}$)              &  Method         & ($S_{cut,2}$)                  & ($\sigma_{max,QU}$)              & ($\sigma_{max,I}$)        \\ \hline
0          &   MAD          &       $6 \sigma_1$        &    SD            &   No Rejection                 &      1 mJy                       &     50 mJy            \\
1          &   MAD          &       $6 \sigma_1$        &    SD            &    $4 \sigma_2$                &      1 mJy                       &     50 mJy            \\
2          &   MAD          &       $6 \sigma_1$        &    SD            &    $3 \sigma_2$                &      1 mJy                       &     50 mJy            \\
3          &   MAD          &       $6 \sigma_1$        &    MAD           &     No Rejection               &      1 mJy                       &     50 mJy            \\
4          &   MAD          &       $6 \sigma_1$        &    MAD           &    $4 \sigma_2$                &      1 mJy                       &     50 mJy            \\
5          &   MAD          &       $6 \sigma_1$        &    MAD           &    $3 \sigma_2$                &      1 mJy                       &     50 mJy            \\
6          &   MAD          &       $6 \sigma_1$        &    SD            &     No Rejection               &      2 mJy                       &     50 mJy            \\
7          &   MAD          &       $6 \sigma_1$        &    SD            &    $4 \sigma_2$                &      2 mJy                       &     50 mJy            \\
8          &   MAD          &       $6 \sigma_1$        &    SD            &    $3 \sigma_2$                &      2 mJy                       &     50 mJy            \\
9          &   MAD          &       $6 \sigma_1$        &    MAD           &    No Rejection                &      2 mJy                       &     50 mJy            \\
10          &  MAD           &      $6 \sigma_1$         &   MAD            &   $4 \sigma_2$                 &     2 mJy                        &    50 mJy            \\
11          &  MAD           &      $6 \sigma_1$         &   MAD            &   $3 \sigma_2$                 &     2 mJy                        &    50 mJy            \\
12          &  MAD           &      $6 \sigma_1$         &   SD             &   No Rejection                 &     $0.5$ mJy                    &    50 mJy            \\
13          &  MAD           &      $6 \sigma_1$         &   SD             &   $4 \sigma_2$                 &     $0.5$ mJy                    &    50 mJy            \\
14          &  MAD           &      $6 \sigma_1$         &   SD             &   $3 \sigma_2$                 &     $0.5$ mJy                    &    50 mJy            \\
15          &  MAD           &      $6 \sigma_1$         &   MAD            &   No Rejection                 &     $0.5$ mJy                    &    50 mJy            \\
16          &  MAD           &      $6 \sigma_1$         &   MAD            &   $4 \sigma_2$                 &     $0.5$ mJy                    &    50 mJy            \\
17          &  MAD           &      $6 \sigma_1$         &   MAD            &   $3 \sigma_2$                 &     $0.5$ mJy                    &    50 mJy            \\
18          &  MAD           &      $6 \sigma_1$         &  SD              &   No Rejection                 &     1 mJy                        &    25 mJy            \\
19          &  MAD           &      $6 \sigma_1$         &  SD              &   No Rejection                 &     1 mJy                        &    25 mJy            \\
20          &  MAD           &      $6 \sigma_1$         &  SD              &   No Rejection                 &     1 mJy                        &    25 mJy            \\
21          &  MAD           &      $6 \sigma_1$         &  SD              &   No Rejection                 &     1 mJy                        &    100 mJy           \\
22          &  MAD           &      $6 \sigma_1$         &  SD              &   No Rejection                 &     1 mJy                        &    100 mJy           \\
23          &  MAD           &      $6 \sigma_1$         &  SD              &   No Rejection                 &     1 mJy                        &    100 mJy           \\
24          &  SD            &      $6 \sigma_1$         &  SD              &   No Rejection                 &     1 mJy                        &    50 mJy            \\
25          &  N/A           &    No Rejection           &  SD              &   No Rejection                 &     1 mJy                        &    50 mJy            \\
26          &  MAD           &      $3 \sigma_1$         &  SD              &   No Rejection                 &     1 mJy                        &    50 mJy            
\end{tabular}
\caption{Data filter tests.}
\end{center}
\label{tbl:systests}
\end{table*}


\begin{table*}[h]
\begin{center}
\begin{tabular}{l l l}
   & Average ON-OFF & Typical Per-Measurement Noise \\ \hline
 $I$ & $-5.1 \pm 1.4 \mK$ & $42 \mK$  \\
 $Q$ & $35 \pm 30. \uK$ & $0.81 \mK$ \\
 $U$ & $26 \pm 34 \uK$ & $0.93 \mK$ \\
 $V$ & $-0.1 \pm 0.3 \mK$ & $10. \mK$ \\ \hline
\end{tabular}
\caption{Final results of $9.65$ GHz full-Stokes polarization
measurements of L1622 and typical per-measurement noise level.}
\end{center}
\label{tbl:finalresults}
\end{table*}

\begin{table*}[h]
\begin{center}
\begin{tabular}{l |c c| c c }
 & \multicolumn{2}{c}{P.A.} & \multicolumn{2}{c}{MJD} \\ 
 & $\chi_{\nu}^2$ & P.T.E.    & $\chi_{\nu}^2$ & P.T.E.\\ \hline
 $I$ & $0.44$ & $78\%$  & $0.74 $ & $56\%$ \\
 $Q$ & $0.98$ & $42\%$ & $1.97$ & $10\%$ \\
 $U$ & $0.52$ & $72\%$ & $0.54$ & $70\%$ \\
 $V$ & $2.44$ & $4\%$ & $1.92$ & $10\%$ \\ \hline
\end{tabular}
\caption{$\chi_{\nu}^2$ values for data binned in Parallactic Angle and
time, with respect to the final fully averaged values, all with $\nu = 4$.}
\label{tbl:binnedchi}
\end{center}
\end{table*}



\section{Results \& Conclusions}

\subsection{Total Intensity Signal}
\label{sec:Stokesiinterp}

We have measured a total intensity signal (On-Off) of $-5.1 \mK$
$(-2.6 \mJyBm)$ towards L1622, indicating that at $9.65 \ghz$ the sky
is brighter off-source than on-source. To understand the nature of
this signal we must consider both spinning dust and free-free, which
will each be significant at $9 \ghz$. Fortunately high-resolution
H$\alpha$ and $100 \um$ data that cover the region of interest are
available.

\begin{figure*}[h!]
\plotone{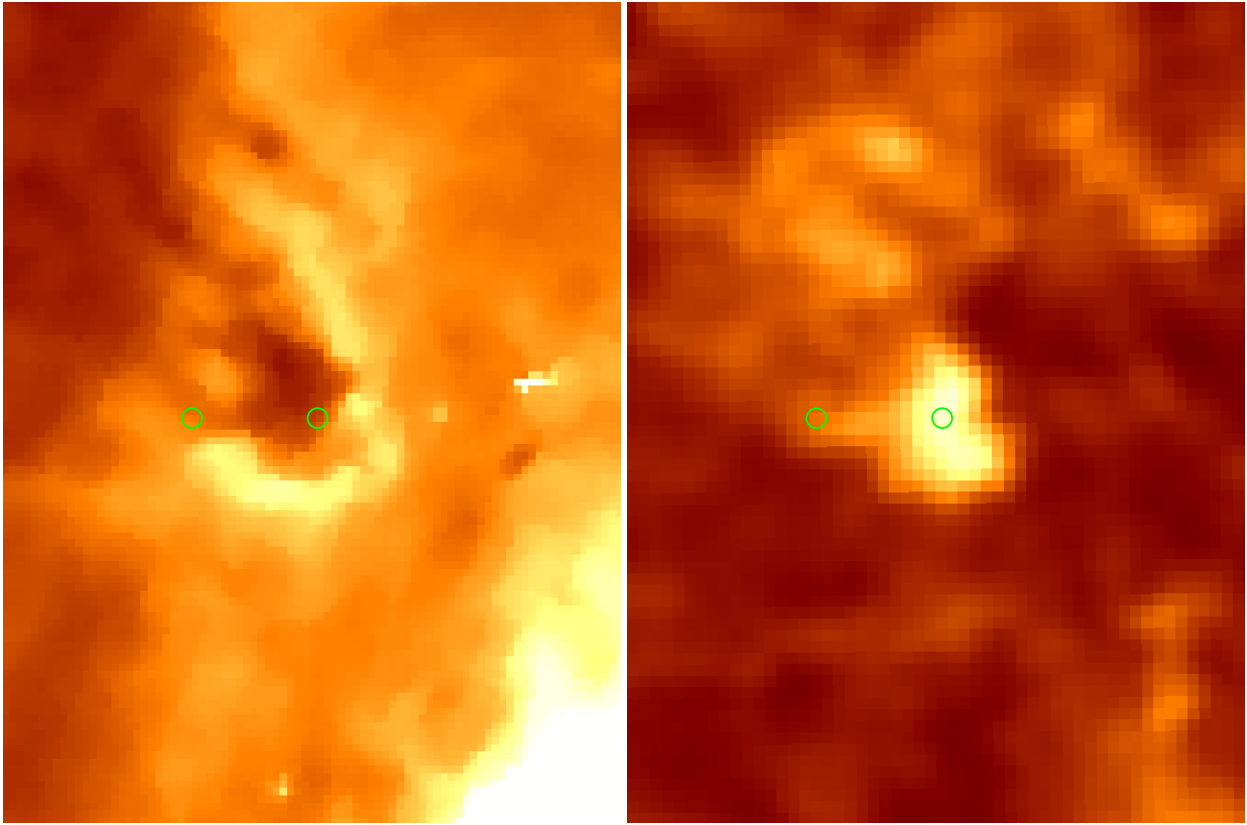}
\caption{H$\alpha$ map (left) and 100 $\mu$m map (right). Our on and
off positions are shown as green circles. }
\label{fig:maps}
\end{figure*}

L1622 is embedded within the Orion star forming complex, in a vicinity
of significant free-free emission. Using the SHASSA map
\citep{gaustad01} we determine that there is a 20 Rayleigh difference
between our on-source and off-source locations. The H$\alpha$
measurements suffer from significant extinction by dust, which must be
accounted for in order to accurately predict the radio free-free
signal.  We estimate the extinction from the \citet{sfd98} maps of
temperature-corrected $100 \mu m$ dust intensity.  The SHASSA and IRAS
$100 \um$ maps are shown in Figure~\ref{fig:maps}, along with our On
and Off source positions. For display purposes we use the more recent
and higher-quality \citet{iris} maps, however, the dust extinction
relations  are based on the \citet{sfd98} $D^T$ maps. In the
SHASSA map L1622 appears as a dark cloud, ostensibly between us and
the majority of the H$\alpha$ emission. The peak dust-corrected
infra-red brightness on-source is $\sim 160 \mJySr$; allowing for a
characteristic background level of $60 \mJySr$ in the vicinity, the
peak emission due to the dark cloud itself is $100 \mJySr$. With an
H$\alpha$ extinction relation of $0.0462 \, {\rm mag/(MJy/Sr)}$
\citep{dickinson03} we find an overall H$\alpha$ extinction of $4.6 \,
{\rm mag}$. At this high level of absorption the free-free predictions
are not reliable.  We can nevertheless estimate the characteristic
level of free-free emission variations nearby, which we find to be
$\pm 5 \mK$  on scales of our beam throw (66 seconds or $16'.5$)
using the $T_e=7000K$ scaling of H$\alpha$ intensity to radio surface
brightness ($60.9 \,{\rm \mu K/Rayleigh}$ at 10 GHz ---
\citet{dickinson03}). 

With the limited frequency coverage of our two-point measurement we
are unable to distinguish directly between free-free and ``anomalous''
excess emission. To do this we can we use the 8 and 10 GHz
measurements of \citet{fink02}, who, in linear scans across L1622
determine a peak dust-correlated excess signal of 4 mK. We can compare
this to the 31 GHz \citet{casassus06} detection of excess anomalous
emission in L1622, which yield a dust emissivity of $24.1 \pm 0.7 \,
{\rm \mu K/(MJy/Sr)}$ (at 31 GHz). Extrapolating to $9.5$ GHz with a
37\% CNM + 63 \% WNM \citep{draine98b} dust SED
\citep{casassus06,fink02} gives a dust emissivity of $54.6 \, {\rm \mu
K/(MJy/Sr)}$, consistent with the value de Oliveira-Costa et
al. (1999) give at 10 GHz, $50 \, {\rm \mu K/(MJy/Sr)}$. These
scalings, together with the extinction-corrected levels of dust
emission in the field, predict a $9.65$ GHz spinning dust signal of 3
mK. We adopt an estimated $9.65$ GHz,excess dust-correlated signal
level in Stokes I of $3.5 \pm 0.5 \mK$.

We note that this measurement and that of \citet{fink02} use the same
central position on L1622 but different reference (off-source)
positions; the chop throws are comparable ($16'$ vs $12'$).

\subsection{Polarization Signal}

With the GBT at 9 GHz we have determined that $Q = 35 \pm 30 \uK$ and
$U = 26 \pm 34 \uK$. Figure~\ref{fig:qulimits} shows the joint
two-parameter $68.3 \%$, $95.4\%$, and $99.7\%$ confidence regions for
fractional $Q$ and $U$.  We employ a simple Maximum Likelihood
approach to set a limit on the total linear polarization that avoids
the Rice bias. The likelihood of a linear polarization $p=\sqrt{Q^2 +
U^2}$ and polarization angle $\theta$ given statistically independent
data $Q_{obs},U_{obs}$ is
\begin{equation}
L(p,\theta|Q_{obs} , U_{obs}) \propto
Exp\left[- \left( \frac{(Q-Q_{obs})^2}{2\sigma_{Q_{obs}}^2} +\frac{(U-U_{obs})^2}{2\sigma_{U_{obs}}^2}  \right)   \right]
\end{equation}
where $Q=p \, cos \, 2\theta$ and $U = p \, sin \, 2\theta$. 
Marginalizing over the position angle of polarization we have
\begin{equation}
L(p|Q_{obs} , U_{obs}) \propto \int \, d\theta  \, 
Exp\left[- \left( \frac{(Q-Q_{obs})^2}{2\sigma_Q^2} +\frac{(U-U_{obs})^2}{2\sigma_U^2}  \right)   \right]
\end{equation}
The PDF is normalized to unity, and confidence intervals are
determined by integrating the PDF.  We find $95.4\%$ and $99.7\&$
upper limits on the total linear polarization $p$ of $88 \uK$ and $123
\uK$, respectively. For a nominal Stokes I dust signal of $3.5 \pm 0.5
\mK$ (\S~\ref{sec:Stokesiinterp}), and assuming that the free-free
signal is unpolarized, these limits correspond to fractional linear
polarizations of $2.7\%$ and $3.5 \%$.

These limits are consistent with the results of Battistelli et al.  on
Perseus, who find a fractional linear polarization of
$3.4^{+1.5}_{-1.9} \%$.  If electric dipole emission is primarily
responsible for the observed microwave excess, a population of small
dust grains is required, with tyipcal radii of a few nm. These grains
can be efficiently aligned by paramagnetic resonance relaxation
(Lazarian \& Draine 2000), which would give rise to up to $\sim 5\%$
fractional linear polarization at 9 GHz.  Both our result and that of
Battistelli et al. are lower than this, but probably within
theoretical uncertainties.  In contrast, magnetic dipole fluctuations
in larger, single-domain ferrous grains will show a fractional linear
polarization $\sim 10\%$ below 10 GHz, flipping orientation and rising
to as much as 33\% at 100 GHz (Draine \& Lazarian 1999). Many of the
magnetic dipole models are excluded by our measurements.


\section{Conclusions}

We have presented a $9.65$ GHz limit on linear polarization towards
the dark cloud Lynds 1622, a well-measured locus of anomalous Galactic
microwave emission; the total degree of linear polarization is $<88
\uK$ at $2\sigma$ ($<123 \uK$ at $3\sigma$). Assuming a $3.5 \mK$
stokes I spinning dust signal, consistent with independent
measurements of L1622 at 8 through 32 GHz \citep{fink02,casassus06},
and that the free-free signal is unpolarized, we have limits on the
degree of linear polarization of $2.7\%$ and $3.5\%$.  These limits
are consistent with the expected linear polarization of small rotating
grains, and with the 3\% fractional linear polarization measured by
Battistelli et al. towards Perseus, but inconsistent with many models
for the anomalous emission based on magnetic dipole fluctuations in
ferrous grains. Such low levels of linear polarization would also be
unusual for soft, dust-correlated synchrotron, which has been
suggested as a possible origin for the generally observed anomalous
microwave emission \citep{bennet03}. For L1622, however, soft
synchrotron was already strongly ruled out by existing total intensity
data.  In addition we see that the total intensity signal on the line
of sight to L1622 is {\it fainter} than on the nearby comparison
(Off-source) region.  The negative signal we observe is consistent
with the level of free-free fluctuations in the region: considering
the expected $+3.5$ mK signal, the observed -5 mK signal indicates an
$8.5$ mK gradient between our ON and OFF positions, consistent with
our estimate of $\pm 5 \mK$ variations over the region from H$\alpha$
maps.  This underscores the need for good frequency coverage in
attempts to charactize the spinning dust foreground at low microwave
frequencies.  The relatively weak polarization signal observed in
L1622 and Perseus bodes well for future CMB polarization
experiments. It should be appreciated that while these concentrated
loci of emission are important first steps on the road to
understanding the anomalous microwave emission, the physical
conditions are quite different from those in the diffuse ISM, where
foreground properties are most important for future CMB
expreiments. To study these regions more sensitive observations are
necessary.

The National Radio Astronomy Observatory is a facility of the National
Science Foundation operated under cooperative agreement by Associated
Universities, Inc.

\bibliography{gbtspdustpolx}

\end{document}